\documentclass[10pt]{article}
\usepackage{array,amsmath,amsfonts,graphicx,mathrsfs}
\usepackage[colorlinks=true]{hyperref}
\usepackage[T1]{fontenc}
\newcommand{\bs}{\boldsymbol}

\newtheorem{theorem}{Theorem}

\begin{document}

\centerline{\Large\bf Random threshold for linear model selection, revisited.}

\vspace{2pt}

\vspace{.4cm}
\centerline{Merlin Keller$^{1},$ Marc Lavielle$^{2},$}
\vspace{.4cm}
\centerline{\it$^1$ EDF R\&D, Chatou, France}
\centerline{\it $^2$ INIRIA Saclay, France and University of Paris Sud, Orsay, France}

\vspace{.55cm}
\fontsize{9}{11.5pt plus.8pt minus .6pt}\selectfont

\tableofcontents

\paragraph{Abstract}

In \cite{Lavielle07}, a random thresholding method is introduced to select the significant, or non null, mean terms among a collection of independent random variables, and applied to the problem of recovering the significant coefficients in non ordered model selection. We introduce a simple modification which removes the dependency of the proposed estimator on a window parameter while maintaining its asymptotic properties. A simulation study suggests that both procedures compare favorably to standard thresholding approaches, such as multiple testing or model-based clustering, in terms of the binary classification risk. An application of the method to the problem of activation detection on functional magnetic resonance imaging (fMRI) data is discussed.

\section{Introduction}

In \cite{Lavielle07}, the following model is considered:
\begin{eqnarray}\label{eq:obs_model}
Y_i &=& \mu_i + \varepsilon_i, \quad i = 1, \ldots, n,
\end{eqnarray}
where $\mu_i$ are unknown constants, some of which are zero, and $\varepsilon_i$ are independent, identically distributed (iid) zero-mean random variables, with known cumulative distribution function (cdf) $F_\varepsilon.$

Within this model, the problem of selecting the significant coefficients $\mu_i \neq 0,$ based on the observations $Y_i$ is studied. Such a problem arises in many different application areas, such as genomics \cite{Ge03}, or neuroimaging \cite{Friston97}, to cite just a few.

Many methods have been proposed to perform this task. Multiple testing procedures for instance (see \cite{Ge03} for an overview of existing methods), have been developed to control a certain type I error rate, such as the familywise error rate (FWER), or the false discovery rate (FDR) \cite{Benjamini95} at a user-fixed level. It can be argued however that the choice of a level, which ultimately defines the subset of selected coefficients, is arbitrary, as their is no safe guideline to what an `optimal' level of false detections should be.

An alternative, that allows to control both type I and type II error rates, consists in fitting a mixture model to the data, with one class for the null (zero-mean) data, and one, or more, for the non-null data. A detection threshold can then be derived, which minimizes a certain classification risk, such as the binary risk, associated to the 0\,-1 loss function, resulting in a `naive Bayes' classifier \cite{MacKay03}. The main difficulty of this approach lies in the choice of a distribution for the non-null data, which may influence significantly the resulting classifier. Many authors have proposed to deal with this issue through model selection techniques (see \cite{Hastie01,Massart03,Efron04} for instance), however it remains an open-ended problem.

In view of these difficulties, the random threshold (RT) approach introduced in \cite{Lavielle07} appears as a promising candidate, since it does not require the specification of a type-I error level, nor of a model for the non-zero mean observations. The principle of RT lies in estimating the number of significant coefficients, based on a random centering of the partial sums of the ordered observations. Because it relies on as little assumptions as possible, we expect RT to be more robust than the above-mentioned approaches.

However, to date very little is known concerning the classification performances of RT procedures; \cite{Lavielle07} essentially gives a minimal separation speed between null and non-null data for the method to attain perfect classification asymptotically. Furthermore, the algorithm described therein still depends on a window parameter, which may have some influence in presence of noisy data.

This article describes a simple modification of the RT procedure, which removes its dependency on the window parameter, while maintaining its asymptotic properties. We then study the classification performances of both techniques using numerical experiments, in comparison to the above-mentioned standard approaches.

The rest of this article is organized as follows. In Section~\ref{sec:original_procedure}, the original RT method is reviewed. The variable window extension is introduced in Section~\ref{sec:extensions}. In Section~\ref{sec:simulation_study} the results of numerical experiments are presented, which show the good properties of RT in terms of classification. An application to fMRI data analysis is discussed in Section~\ref{sec:fMRI}, and we conclude in Section~\ref{sec:conclusion} by considering the perspectives open by this work.

\section{Original random thresholding procedure}\label{sec:original_procedure}


\subsection{Testing the presence of significant coefficients}\label{sec:test}

\begin{figure}
\centering
\includegraphics[width=\textwidth]{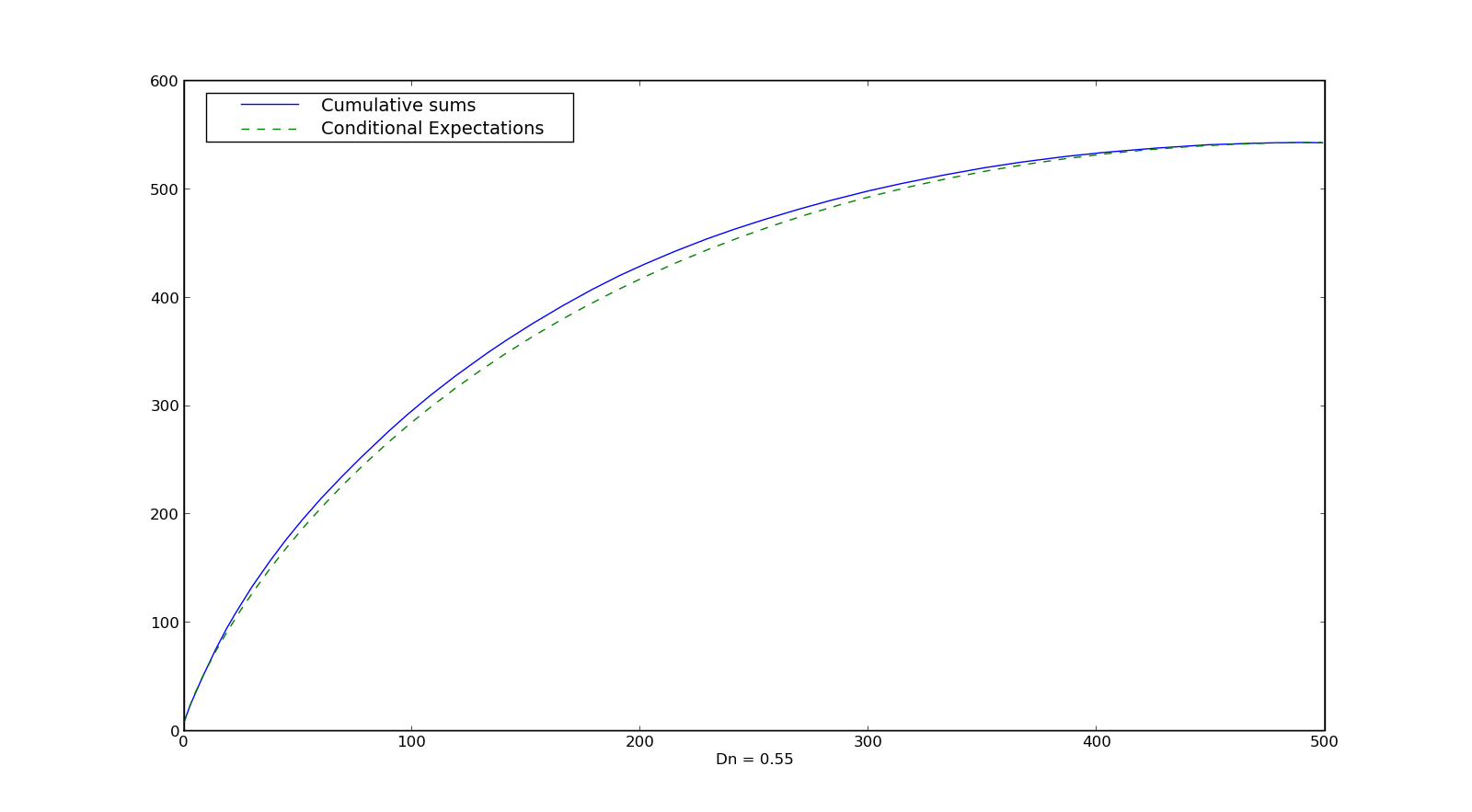}\\
\includegraphics[width=\textwidth]{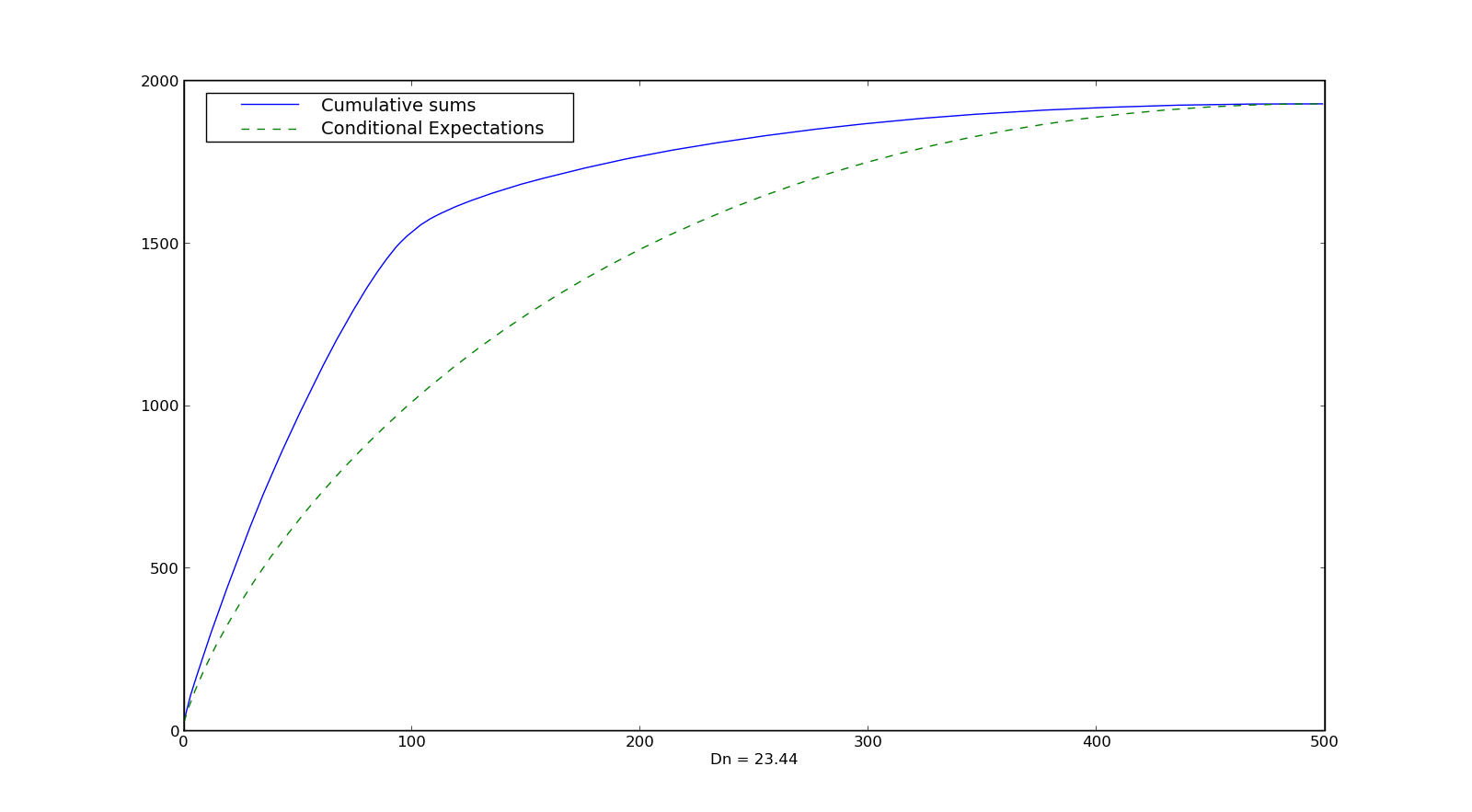}\\
\caption{\label{fig:test_statistic} Test of the presence of significant coefficients: the cumulative sums $(T_j)$ and their conditional null expectations $(Q_j),$ under the global null $\mathcal H_0$ (top) and in presence of significant coefficients (bottom).}
\end{figure}

We start by recalling how the presence of non-zero means is tested, that is, how the null hypothesis $\mathcal H_0: \mu_i \equiv 0$ is tested in \cite{Lavielle07}. This is done by comparing the cumulative sums of the ordered observations to their conditional expectations under $\mathcal H_0,$ according to the following steps:

\begin{enumerate}
\item Order the observations $|Y_{(1)}| \geq |Y_{(2)}| \geq \ldots \geq |Y_{(n)}|$
\item For $i = 1, \ldots, n$,\quad let $X_{(i)} = - \log(1 - F_{|\epsilon|}(|Y_{(i)}|))$
\item Let $T_j = \sum_{i= 1}^j X_{(i)}$ and $Q_j = \mathbb E_{\mathcal H_0} (T_j | T_n)$
\item Define the test statistic $D_n = \max_j |T_j - Q_j| / \sqrt{n}$. The null hypothesis is rejected if $D_n > d_{\alpha}$, with $d_{\alpha}$ such that $\mathbb P_{\mathcal H_0}(D_n > d_{\alpha}) \leq \alpha.$
\end{enumerate}

Note that the cumulative sums are not computed directly from the ordered observations, but from the transforms $X_{(1)}, \ldots, X_{(n)}$ which, under $\mathcal H_0,$ are an ordered series of $\mathcal E(1)$ random variables. The conditional expectations $Q_j = \mathbb{E}_{\mathcal H_0} (T_j | T_n)$ can then be computed using the following result (see \cite{Lavielle07} for details);

\begin{theorem}[Conditional expectations of ordered exponential variables]\label{theo:cond_mean}
Under $\mathcal H_0,$ the $X_{(i)}$ are an ordered series of $\mathcal E(1)$ random variables. It follows that:
\begin{itemize}
\item[i)] $\mathbb{E}_{\mathcal H_0} (X_{(i)}) = \sum_{\ell=i}^n \frac{1}{\ell}$
\item[ii)] $\mathbb{E}_{\mathcal H_0} (T_j) = j + j \sum_{\ell=j}^n \frac{1}{\ell} $
\item[iii)] $\mathbb{E}_{\mathcal H_0} (T_j | T_n) = \frac{\mathbb{E}_{\mathcal H_0} (T_j)}{\mathbb{E}_{\mathcal H_0} (T_n)} T_n.$
\end{itemize}
\end{theorem}

Furthermore, for $n \geq 100,$ it is shown in \cite{Lavielle07} using Monte-Carlo simulations that:
\begin{eqnarray}\label{eq:approx_calibration}
\mathbb P_{\mathcal H_0}(D_n > 0.65) &\approx& 0.05,
\end{eqnarray}
which provides an approximate calibration for the above test. It is illustrated in Figure~\ref{fig:test_statistic}, on a dataset of $n=500$ observations $Y_i$ simulated according to model~(\ref{eq:obs_model}), with noise terms $\varepsilon_i$ sampled from the $\mathcal N(0,1)$ distribution.

Under the global null $\mathcal H_0,$ that is, when $\mu_i \equiv 0,$ (Figure~\ref{fig:test_statistic}, top), the cumulative sums $(T_j)$ are seen to follow closely their conditional expectations $(Q_j).$ Consequently, the resulting test statistic value $D_n = 0.55,$ does not exceeds the critical value $0.65$ given by (\ref{eq:approx_calibration}).

In contrast, when we add $n_1=100$ non-zero means, all taken equal to $\mu_i \equiv 5.,$ to the null data (Figure~\ref{fig:test_statistic}, bottom), the $T_j'$s become substantially larger than their expected values $Q_j$ under $\mathcal H_0,$ resulting in a gap between the corresponding curves. Note that this gap is most significant around $j=100,$ because, for all $j,$ $T_j$ is the sum of the $j$ largest observations (after transformation), containing mostly non-zero means for $j\leq 100.$ Consequently, the ensuing test statistic $D_n = 23.44$ is far above the critical value $0.65.$

\subsection{Selecting the significant coefficients}\label{sec:procedure}

\begin{figure}
\centering
\includegraphics[width=0.9\textwidth]{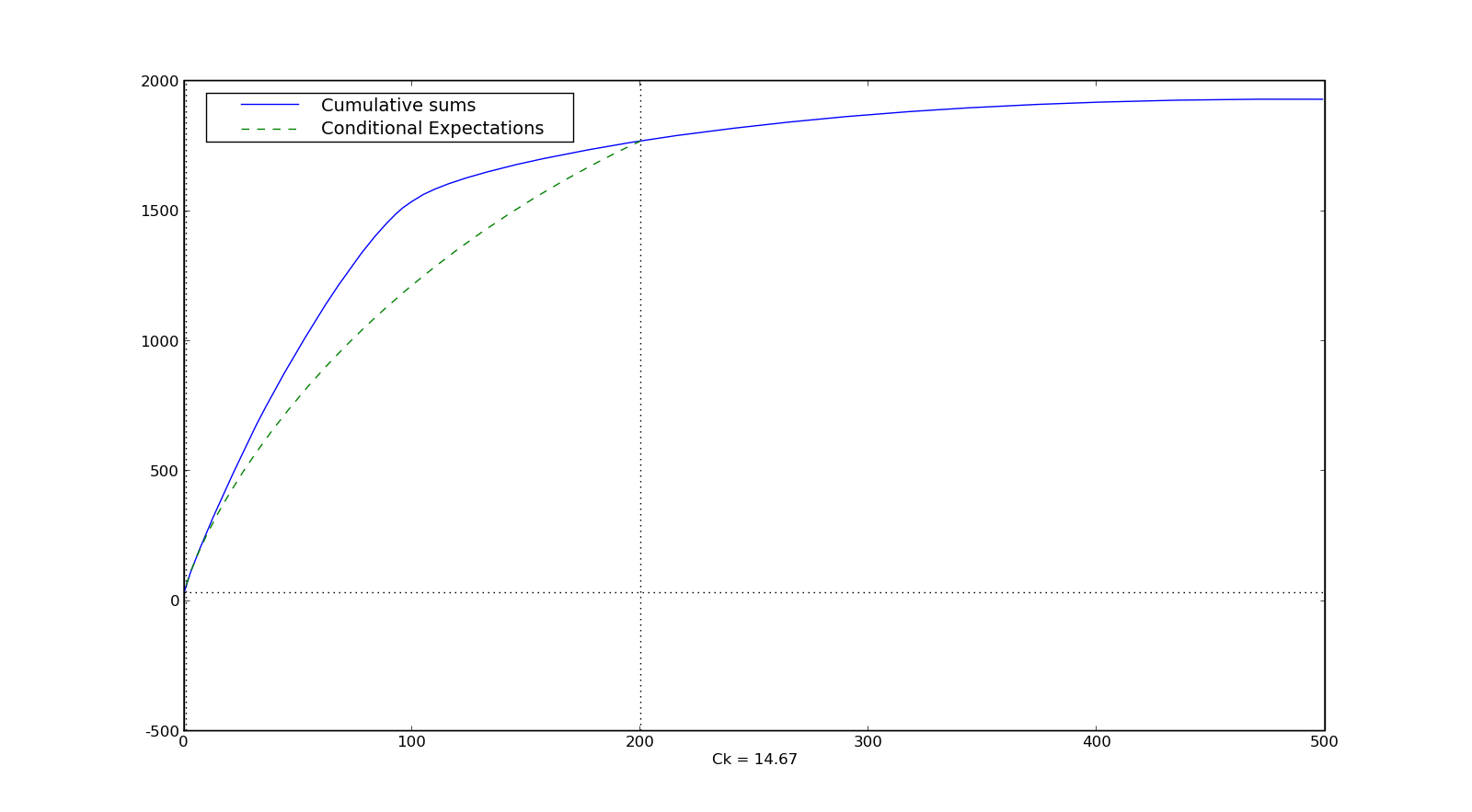}\\
\includegraphics[width=0.9\textwidth]{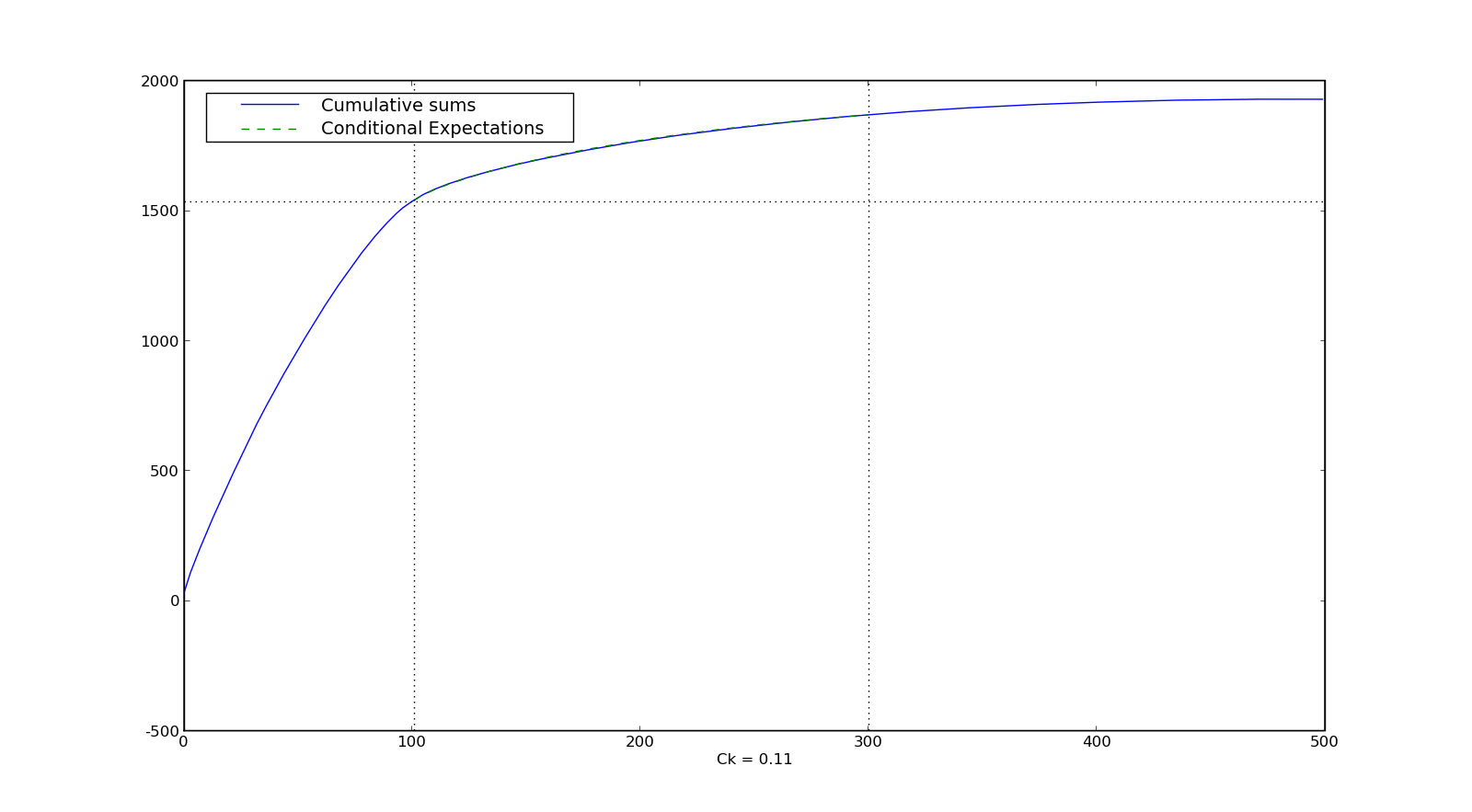}\\
\includegraphics[width=0.9\textwidth]{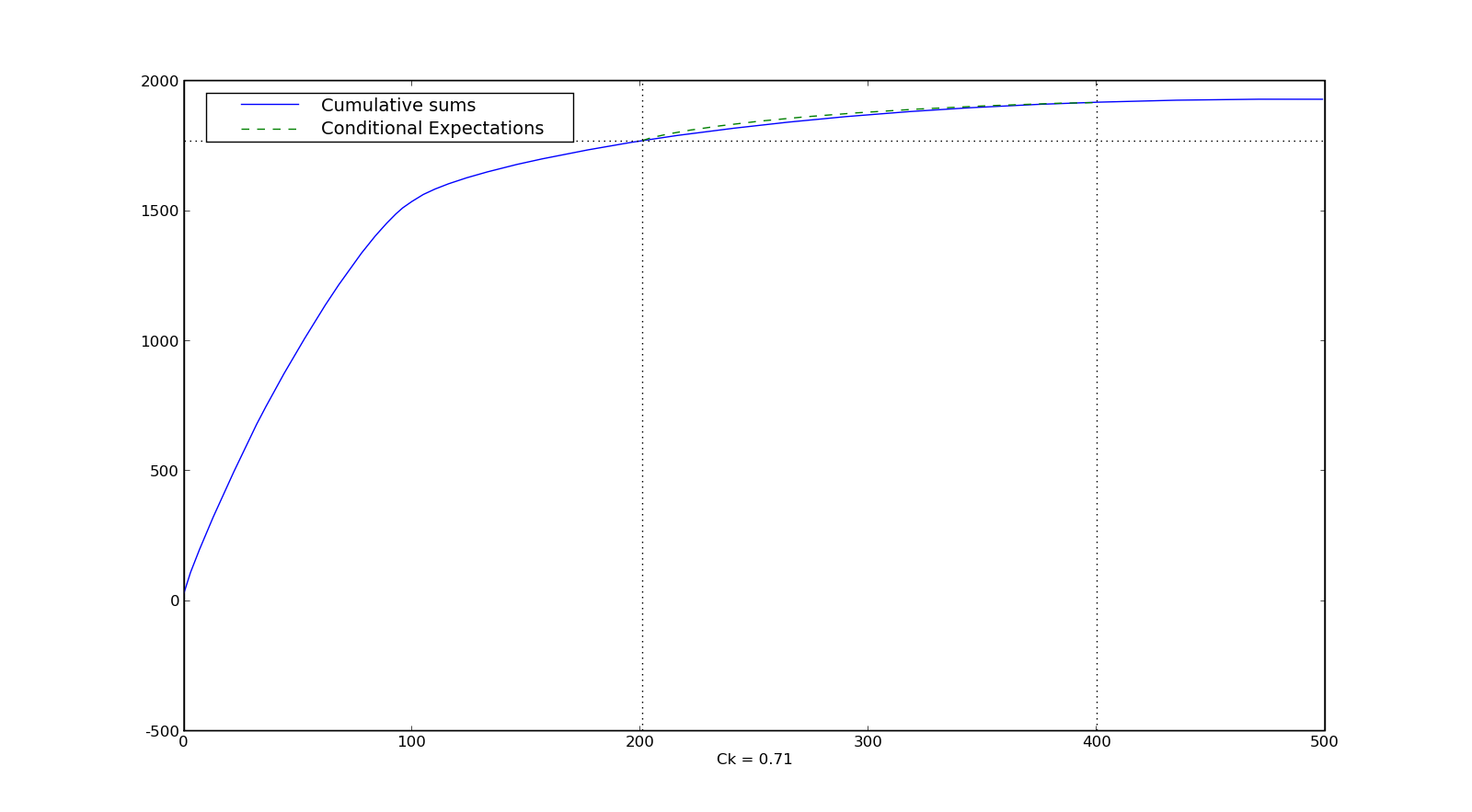}\\
\caption{\label{fig:RT_examples} Random threshold procedure: partial cumulative sums $(T_{k,j})_j$ and their conditional null expectations $(Q_{k,j})_j$ for $k = 0$ (top), $k = 100$ (middle), and $k = 200$ (bottom), with window width $K_n = 200.$}
\end{figure}

\begin{figure}
\centering
\includegraphics[width=\textwidth]{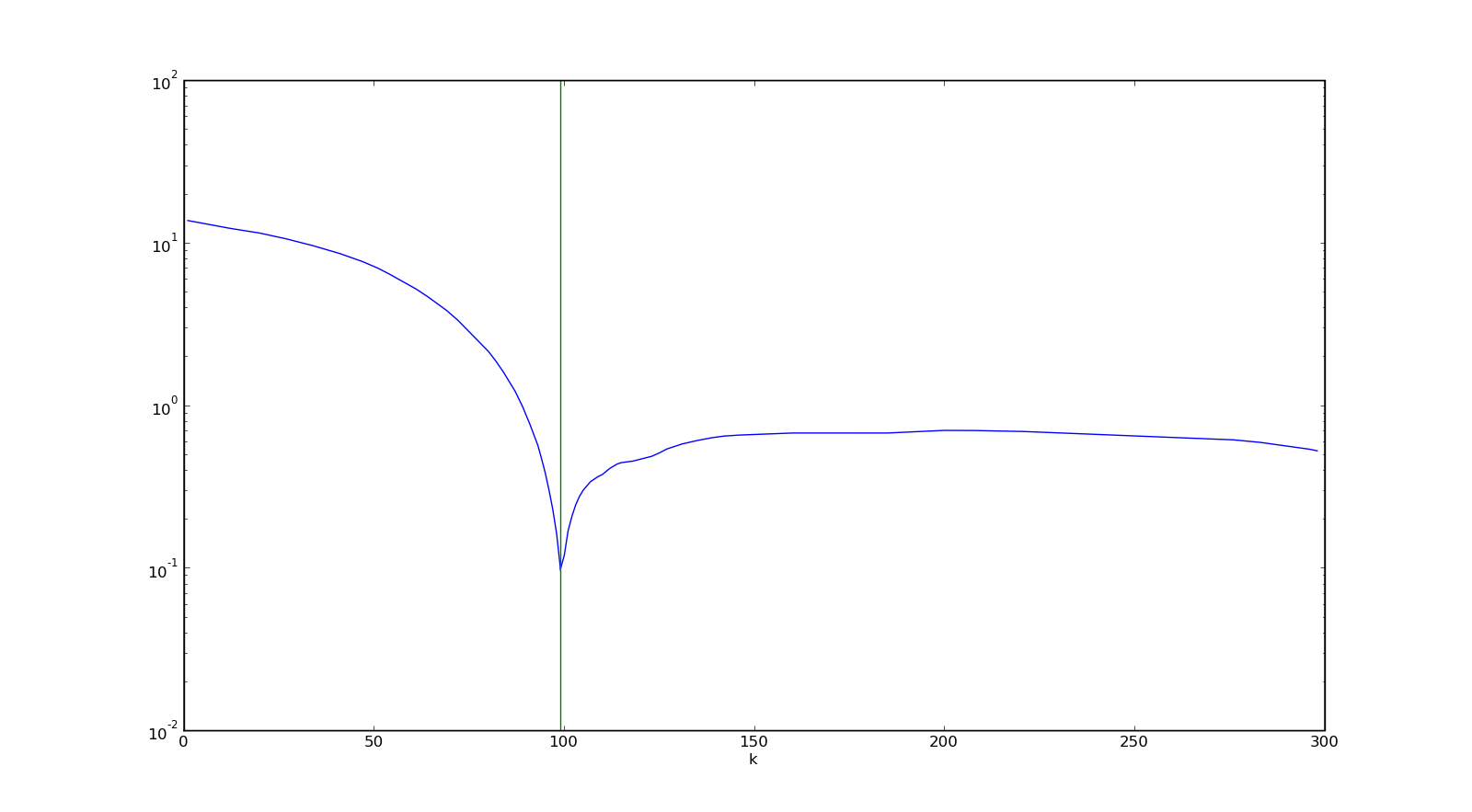}
\caption{\label{fig:criterion} Random threshold procedure: the sequence $\eta_k$ on a logarithmic scale; its minimum is attained for $\hat k_n=99.$}
\end{figure}

Upon rejection of the null hypothesis, the following task consists in selecting the significant coefficients. The procedure for doing so can be interpreted in a data-dependent `multiple hypothesis testing' setting, as described hereafter. Consider the null hypothesis $\mathcal H_0,$ as defined in Section~\ref{sec:test}, and the set of alternative hypotheses:

\begin{itemize}
\item[] $\mathcal H_{1,k} :$ for any $i \leq k,\ \mu_{(i)} > 0,$ and $\mu_{(k+1)} = \ldots = \mu_{(n)} = 0.$
\end{itemize}

In other terms, $\mathcal H_{1,k}$ corresponds to the hypothesis that the $k$ largest observations only have non-zero means. Even though in real-life datasets null and non-null data are never perfectly separated, in general one cannot expect more than to discriminate between such hypotheses in non-ordered model selection. Note that this is equivalent to choosing a certain detection threshold to separate null from non-null data.

Denote $\mathbb{E}_k$ the expectation under $\mathcal H_{1,k}$ (instead of $\mathbb{E}_{\mathcal H_{1,k}}$). The RT procedure first computes the $X_{(i)}$'s using the same steps 1. and 2. as in Section~\ref{sec:test}, then adds the following steps:

\begin{itemize}

\item[3.] Let $K_n$ be some positive integer. For $1 \leq k \leq n - K_n$ and $1 \leq j \leq K_n,$ compute:
\begin{eqnarray}
T_{k,j} & = & \sum_{i = k + 1}^{k + j} X_{(i)} \nonumber\\
Q_{k,j} & = & \mathbb{E}_k (T_{k,j} | T_{k,K_n}) \nonumber\\
\eta_k & = & \max_{1 \leq j \leq K_n} |T_{k,j} - Q_{k,j}| / \nonumber\sqrt{n}.
\end{eqnarray}

\item[4.] Let $\hat k_n = \mathrm{argmin}_{1 \leq k \leq n - K_n} \eta_k.$

\end{itemize}

As in the preceeding section, for each $k$ $X_{(k + 1)}, \ldots, X_{(k + n - k)}$ is an ordered series of $\mathcal E(1)$ variables under $\mathcal H_{1,k},$ so $Q_{k,j}$ can easily be computed using Theorem~\ref{theo:cond_mean} (see \cite{Lavielle07} for further details). Heuristically, the partial cumulative sums $(T_{k,j})$ are compared to their conditional expected values $Q_{k,j}$ under the hypotheses $\mathcal H_{1,k},$ for $k = 1, \ldots, n - K_n.$ The number of significant coefficients is estimated as the index $\hat k_n$ corresponding to the minimal gap between $T_{k,j}$ and $Q_{k,j},$ as evaluated by $\eta_k.$

This is illustrated in Figure~\ref{fig:RT_examples}, using a a dataset simulated exactly as in Section~\ref{sec:test}, that is, with $n_1 = 100$ significant coefficients. Informally, the procedure uses a sliding window with width $K_n,$ and compares the cumulative sums $(T_j)$ within this window to their conditional expectations under the hypothesis that the window contains the $K_n$ largest null terms. For $k=1,$ the window contains in fact mainly significant terms, so that the $T_{1,j}'s$ are well above their expected values, yielding a normalized gap of $\eta_1 = 14.67.$ For $k=100,$ the window indeed contains mostly the $K_n$ largest null terms, so $T_{100,j}$ and $Q_{100,j}$ are of the same order, yielding a much smaller gap $(\eta_{100} = 0.11).$ Finally, for $k = 200,$ the window contains null terms, but not the $K_n$ largest, so the cumulative sums $(T_{k,j})$ become lower than their expected values. Consequently, the gap increases $(\eta_{200} = 0.71).$ Figure~\ref{fig:criterion} shows the complete sequence of $\eta_k$ values, with a clear minimum at $\hat k_n=99,$ close to the true number of significant coefficients.

\section{Extensions and asymptotic properties}\label{sec:extensions}

\subsection{Unknown distribution extension}\label{sec:unknown}

\begin{figure}
\includegraphics[width=\textwidth]{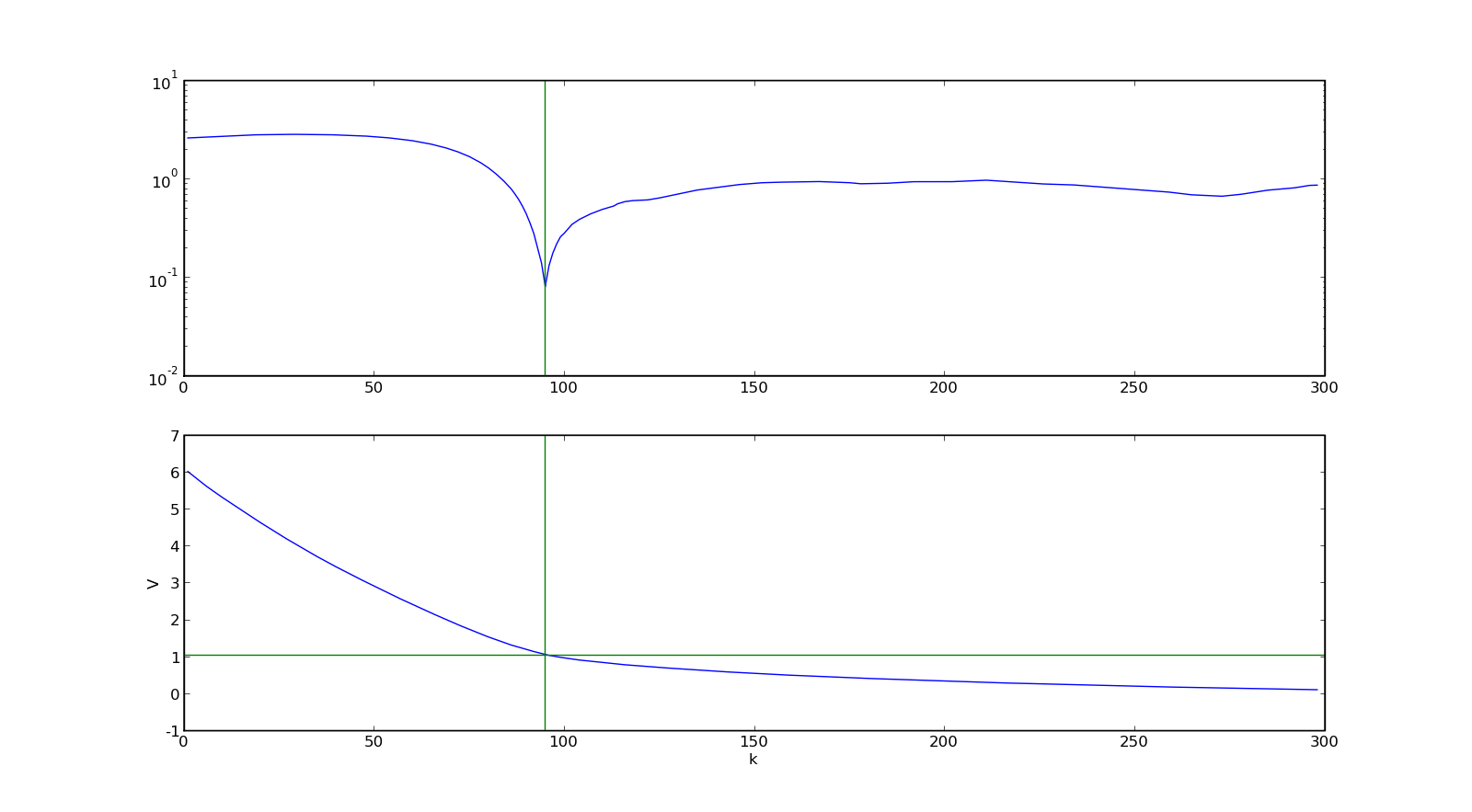}
\caption{\label{fig:gaussian_null} Random threshold method for a Gaussian null distribution with unknown variance $\sigma^2$: the sequence $\eta_k(\hat \sigma_k)$ (top) and the sequence $\hat \sigma_k$ (bottom).}
\end{figure}

The RT method recalled in the previous Sections may be difficult to apply to real-life problems, where the noise distribution $F_\epsilon$ is in general unknown. In \cite{Lavielle07}, an extension is proposed to the case where $F_{\epsilon}$ is a parametric distribution $F_{\epsilon}(\cdot\, ;\, \theta),$ with $\theta$ unknown. Quite naturally, this consists in estimating $\theta$ under each hypothesis $\mathcal H_{1,k}$ from the null data $Y_{(k+1)}, Y_{(k+2)}, \ldots, Y_{(n)},$ then using this estimate to derive the transforms $X_{(i)},$ for $i = k+1, \ldots, k + K_n.$  More precisely, having chosen some positive integer $K_n,$ the extension consists in performing for $1 \leq k \leq n - K_n$ the following steps:
\begin{enumerate}
\item let $\hat{\theta}_k = \hat{\theta}(Y_{k+1}, \ldots, Y_n)$ be an estimate of $\theta$
\item for $i = 1, \ldots, n,$ let $X_{(i)}(\hat{\theta}_k) = -\log \left( 1 - F_{|\epsilon|}(|Y_{(i)}|\, ;\, \hat{\theta}_k) \right),$
\item for $1 \leq j \leq K_n,$ compute:
\begin{itemize}
\item $\quad T_{k,j}(\hat{\theta}_k) = \sum_{i = k + 1}^{k + j} X_{(i)}(\hat{\theta}_k) $\\
\item $\quad Q_{k,j}(\hat{\theta}_k) = \mathbb{E}_k (T_{k,j}(\hat{\theta}_k) | T_{k,K_n}(\hat{\theta}_k)) $\\
\end{itemize}
\item Compute $ \eta_k(\hat{\theta}_k) = \max_{1 \leq j \leq K_n} |T_{k,j}(\hat{\theta}_k) - Q_{k,j}(\hat{\theta}_k)| / \sqrt{n}.$
\end{enumerate}

Finally, the estimated number of components is given as before by 
$$
\hat k_n = \mathrm{argmin}_{1 \leq k \leq n - K_n} \eta_k(\hat{\theta}_k).
$$ 
This simple extension is much more computation intensive than the original procedure, since the $X_{(i)}$'s for $i = k + 1, \ldots, k + K_n$ must be re-computed for each $k,$ instead of once and for all. It is illustrated in Figure~\ref{fig:gaussian_null}, using the same simulated dataset as in the previous Sections. Here the null distribution is defined as the Gaussian $\mathcal N(0; \sigma^2),$ and the unknown variance $\sigma^2$ is estimated by the usual mean squares: $\hat \sigma_k^2 = \frac{1}{n - k} \sum_{i = k + 1}^n Y_{(i)}^2.$ The estimated number of significant coefficient, $\hat k_n = 95$ is still close to its true value, and so is the corresponding standard error estimate $\hat \sigma^2_{\hat k_n} = 1.05.$

\subsection{Varying window extension}\label{sec:varying}

\begin{figure}
\centering
\includegraphics[width=\textwidth]{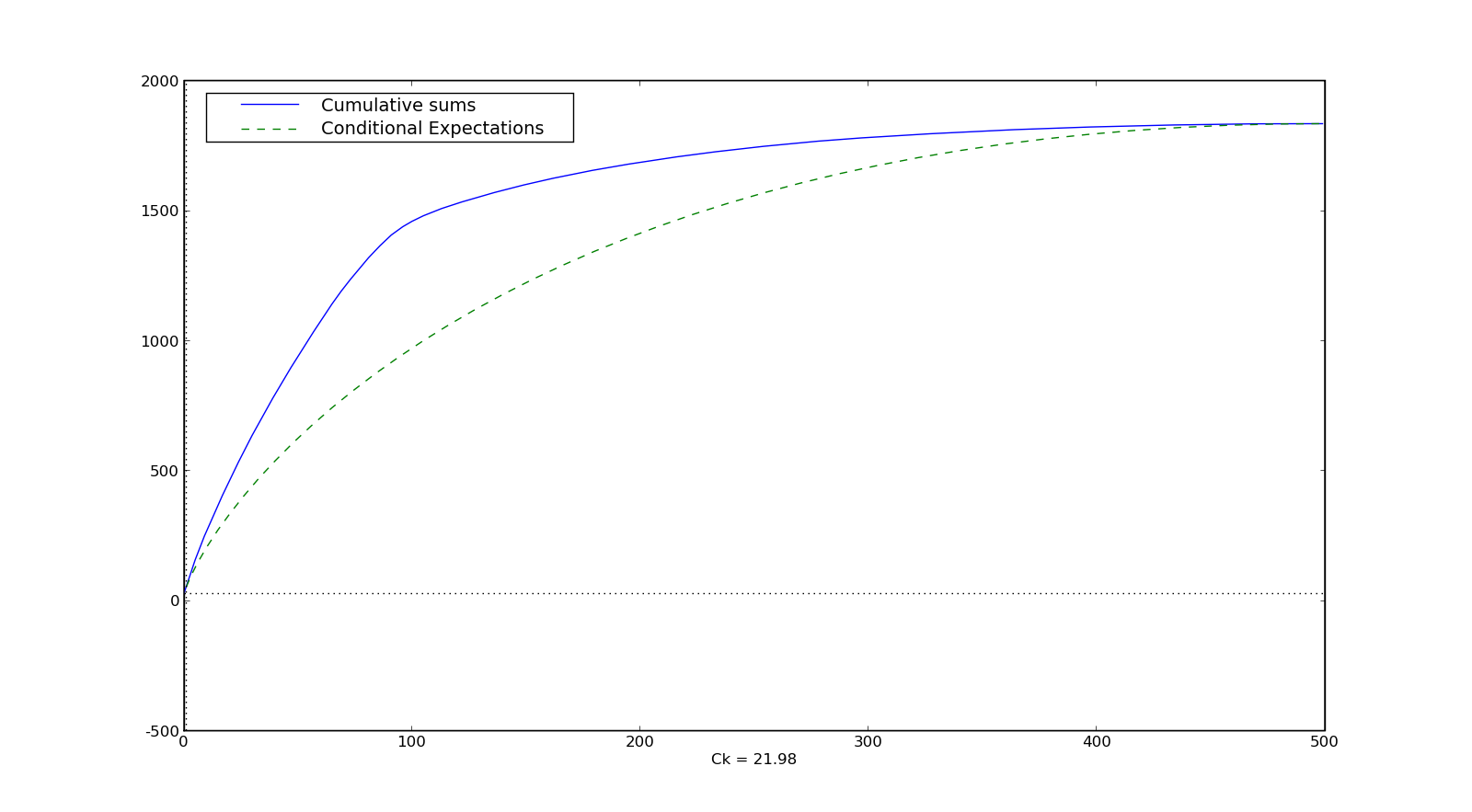}\\
\includegraphics[width=\textwidth]{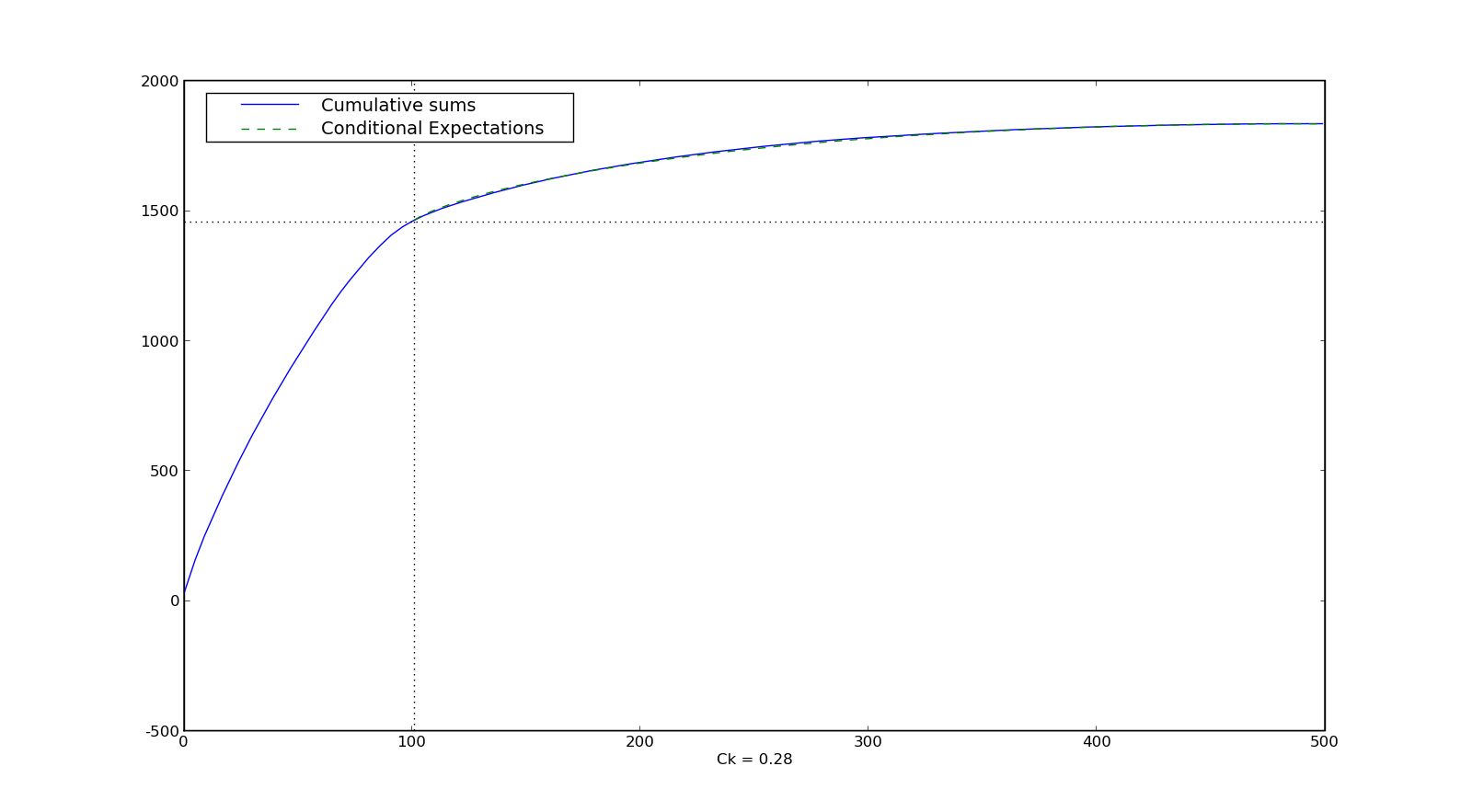}\\
\includegraphics[width=\textwidth]{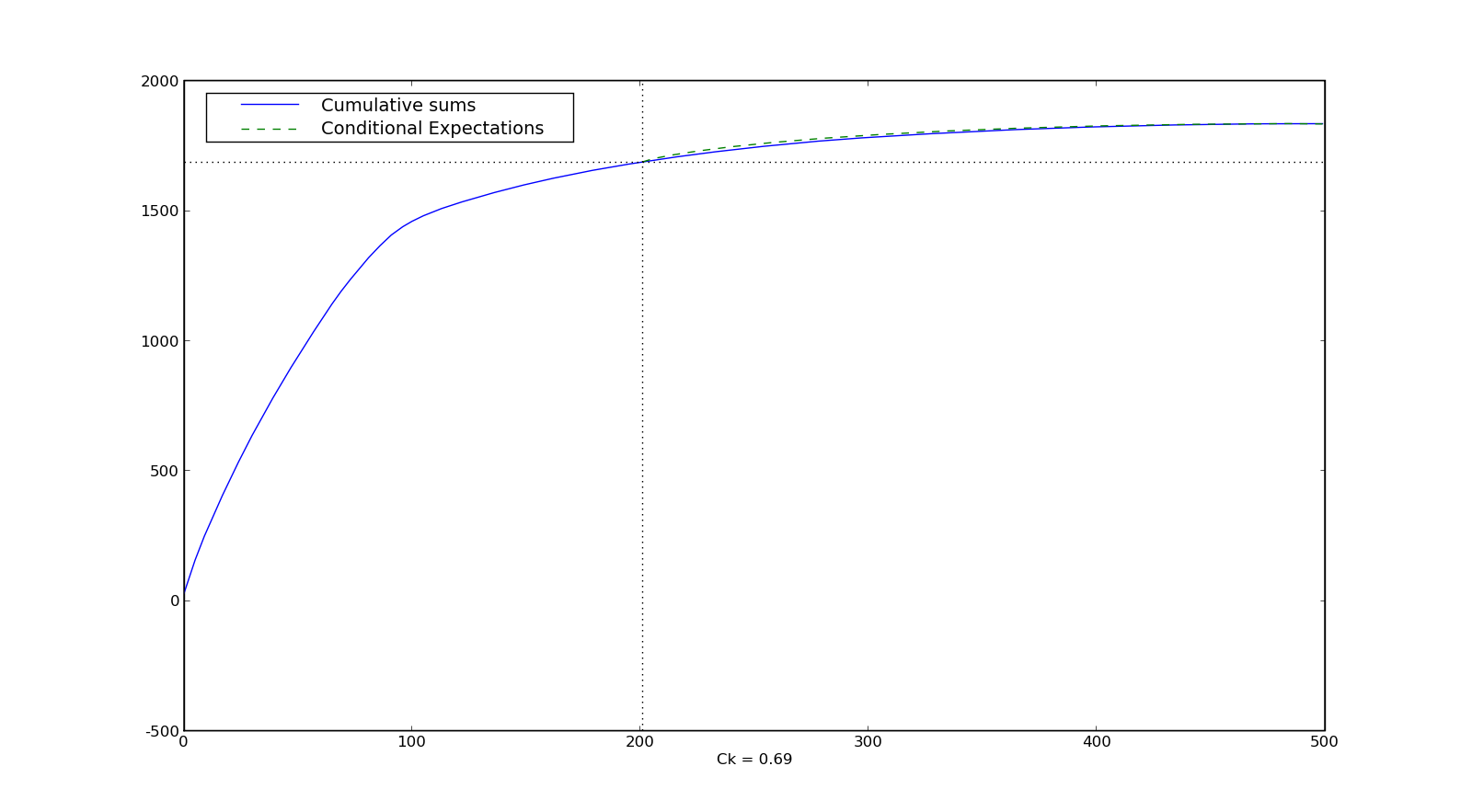}\\
\caption{\label{fig:varwin} Random threshold with a varying window width: partial cumulative sums $(T_{k,j})_j$ and their conditional null expectations $(Q_{k,j})_j$ for $k = 0$ (top), $k = 100$ (middle), and $k = 200$ (bottom).}
\end{figure}

As we have seen previously, the RT procedure depends on a parameter $K_n$ which can be interpreted as a window width, since $\eta_k$ is a function of $X_{(k + 1)}, \ldots, X_{(k + K_n)}.$ $K_n$ must be smaller than the number of null coefficients, but at the same time not too small, or $\eta_k$ would become unstable. Hence choosing an appropriate value for $K_n$ may be a hindrance in practice, especially since we want the RT method to be adaptive and depend as little as possible on any form of tuning.

This issue can be avoided by re-defining $\eta_k$ as a function of $X_{(k + 1)}, \ldots, X_{(n)},$ thus replacing the fixed width $K_n$ by a varying width $n - k,$ which requires no prior tuning. We define the following procedure, starting with the same steps 1. and 2. as in Section~\ref{sec:test}, and adding the following steps:

\begin{itemize}

\item[3.] Let $\kappa_n$ be a lower bound on the number of null coefficients. For $1 \leq k \leq n - \kappa_n$ and $1 \leq j \leq n - k,$ compute:
\begin{equation}\label{eq:procedure}
\begin{array}{ccc}
T_{k,j} & = & \sum_{i = k + 1}^{k + j} X_{(i)} \\
Q_{k,j} & = & \mathbb{E}_k (T_{k,j} | T_{k,n - k}) \\
\eta_k & = & \max_{1 \leq j \leq n - k} |T_{k,j} - Q_{k,j}| / \sqrt{n - k}.
\end{array}
\end{equation}

\item[4.] Let $\hat{k}_n = \mathrm{argmin}_{1 \leq k \leq n - \kappa_n} \eta_k.$

\end{itemize}

In other terms, $\eta_k$ would be strictly equal to the test statistic $D_n$ defined in Section~\ref{sec:test}, if the sequence $(X_{(i)})_{1 \leq i \leq n}$ where replaced by the subsequence $(X_{(i)})_{k + 1 \leq i \leq n},$ \emph{i.e.}, the set of null terms under $\mathcal H_{1,k}.$

Notice that $\hat{k}_n$ is independent from $\kappa_n,$ as long as $\eta$ reaches its global minimum on $\{1, \ldots, n - \kappa_n\}.$ 
The varying window extension presented here can of course be combined with the unknown distribution extension in Section~\ref{sec:unknown}.

Figure~\ref{fig:varwin} illustrates the varying window extension on the same simulated dataset as previously. Intuitively, the `sliding' window is replaced by a `shrinking' window, which at first encloses all observations, than progressively reduces in width as the largest observations are left out. Otherwise the same observations as in the fixed-width case hold: when $k=1,$ the cumulative sums $(T_{k,j})_j$ are larger than their conditional expectations $(Q_{k,j})_j,$ resulting in a large gap ($\eta_0 = 21.98$). This gap is considerably reduced when $k = 100$ as the $T_{k,j}'$s and the $Q_{k,j}'$s become of the same order ($\eta_{100} = 0.28$), then the gap increases again for $k = 200$ as the cumulative sums become smaller than their expected values ($\eta_{200} = 0.69$). Though not shown here, the sequence $\eta_k$ attains its minimum in $\hat k_n = 99,$ as in the fixed-window case.

\subsection{Asymptotic properties}

The estimator of the number of significant coefficients presented in Section~\ref{sec:procedure} is consistent. This is the main result in \cite{Lavielle07}, and it can be extended to the varying window setting. We start by recalling the following asymptotic framework:

\begin{itemize}
\item[{\bf AF1}] There exists $t^{\star} \in (0, 1)$ and a subset $I_{k_n^{\star}}$ of $\{1, \ldots, n\},$ with $k_n^{\star} = [t^{\star} n]$ and $|I_{k_n^{\star}}| = k_n^{\star},$ such that $\mu_i \neq 0$ if $i \in I_{k_n^{\star}}.$ For all other index, $\mu_i = 0.$
\item[{\bf AF2}] For any $i \in I_{k_n^{\star}},$ $|\mu_i| \geq \alpha_n,$ for a certain sequence $(\alpha_n)$ with $\alpha_n \to \infty$ (see \cite{Lavielle07} for further details).
\item[{\bf AF3}] $\kappa_n / n \to c$ such that $0 < c < 1 - t^{\star}.$
\end{itemize}

We then have the following result:

\begin{theorem}[Consistency of the random threshold]\label{theo:consistency}

Let $\hat k_n$ stand for the estimator defined in Section~\ref{sec:procedure}. Under assumptions {\bf AF1}, {\bf AF2}, {\bf AF3}, and appropriate Von-Mises type conditions on the cdf $F_{\epsilon}$ of the errors (see \cite{Lavielle07} for further details), $\hat k_n$ is consistent in the sense that:

\begin{equation}
\label{eq:consistency}
P \bigg( \bigg| \frac{\hat k_n}{n} - t^{\star} \bigg| > u_n \bigg) \to 0,
\end{equation}

for any positive decreasing sequence $(u_n)$ such that $\sqrt{n} u_n \to \infty.$

\end{theorem}

This result can be refined by deriving an upper bound, which we do not detail here, on the convergence rate of the probability in Equation~(\ref{eq:consistency}), for a particular choice of sequence $(u_n).$ Consistency also holds in the unknown distribution case, under a different set of assumptions which we do not recall here, and under the varying window extension, as shown in Appendix~\ref{app:proof}.

This theorem is interesting in that it gives a convergence rate for $\hat k_n,$ provided that a minimal signal-to-noise (SNR) ratio is attained, represented by a lower bound $\alpha_n$ on the absolute values $|\mu_i|$ of the non-zero means (assumption {\bf AF2}). Note that, in order for the random threshold (or any other threshold for that matter) to asymptotically separate perfectly null from non-null data, this SNR must necessarily become arbitrarily large as the sample size increases.

However, this theorem provides no clue to what happens when the SNR remains bounded, as we expect to be the case in real-life applications. In the remainder of this paper, our goal is to explore the behaviour of the RT approach in such cases.

\section{Simulation study}\label{sec:simulation_study}

In order to assess empirically the classification properties of RT, we designed several numerical experiments. Our goal was to compare the binary classification risk of the RT procedure (with both fixed and varying window width), to those obtained by model-based clustering and FDR control techniques. Specifically, we used a mixture-model, estimated via an expectation-maximisation (EM) algorithm \cite{Dempster77} to approximate the risk minimizing detection threshold, and the Benjamini-Hochberg (BH) procedure \cite{Benjamini95} to derive a threshold controlling the FDR at a certain level. We dismissed FWER control techniques as they essentially yield constant thresholds at a given level, and are therefore of little interest when compared to adaptive approaches.

We considered two cases, depending on whether the null distribution $F_\epsilon$ was considered as known or not. Note that the BH procedure is based on the p-values $p_i = 1 - F_{|\epsilon|} (|Y_i|),$ hence it requires that $F_{\epsilon}$ be known, whereas this same distribution can be estimated using the EM algorithm. So in order to compare methods on a fair basis, we compared RT to the BH procedure when the null distribution was known, and to mixture model fit otherwise.

\subsection{Results with known null distribution}

We chose to simulate the $X_i$ directly in the case of a known null distribution. Datasets of $n=10\,000$ observations where generated, containing each $n_1 = 1\,000$ significant terms. These were sampled from the Gamma distribution $\mathcal G(\alpha, \beta),$ where $\beta$ is a scale parameter, and the remaining $9\,000$ null terms were sampled from the $\mathcal E(1)$ distribution.

Table~\ref{tab:FDR_RT_risk} shows the average classification risks obtained by the different methods over $100$ simulated datasets and for different choices of the Gamma distribution parameters. More precisely, we chose to compute the ratio of each attained risk to the lowest achievable (oracle) risk, which makes more sense since perfect classification is in general unattainable.

The binary risk and oracle threshold can be computed as follows. Consider a given dataset $(X_i, Z_i)_i,$ where $Z_i$ is a binary variable, equal to $0$ if $X_i$ is a null term (sampled from the $\mathcal E(1)$ distribution), and $1$ if $X_i$ is a non-null term (sampled from the Gamma distribution). Then the overall classification error associated to a given detection threshold $t$ is given by:
\begin{eqnarray}
c(t) &=& \sum_{Z_i = 0}\bs 1_{\{X_i > t\}} + \sum_{Z_i = 1}\bs 1_{\{X_i <= t\}},
\end{eqnarray}
that is, the sum of type I (false detections) and type II (non detections) errors. The oracle threshold is then chosen to minimize this classification error:

\begin{table}
\centering
\begin{tabular}{ll|lll|lll|lll}
\multicolumn{2}{c|}{} & \multicolumn{3}{c|}{$\beta$} 
& \multicolumn{3}{c|}{$\beta$} & \multicolumn{3}{c}{$\beta$} \\
\multicolumn{2}{c|}{} & 1.0 & 2.0 & 3.0 
& 1.0 & 2.0 & 3.0 & 1.0 & 2.0 & 3.0 \\
\hline\hline
 & $5.0$ & 1.88 & 1.85 & 1.36 & 1.6 & 1.08 & 1.12 & 1.31 & 1.06 & 1.65 \\
$\alpha$ & $6.0$ & 2.4 & 1.66 & 1.14 & 1.59 & 1.04 & 1.61 & 1.19 & 1.33 & 3.00 \\
 & $7.0$ & {\bf 2.7} & 1.37 & 1.09 & 1.42 & 1.21 & {\bf 2.91} & 1.07 & 2.01 & {\bf 6.02}\\
\hline
\multicolumn{2}{c|}{} & \multicolumn{3}{c|}{FDR $0.01$}
& \multicolumn{3}{c|}{FDR $0.05$} & \multicolumn{3}{c}{FDR $0.1$} \\
\end{tabular}
\vskip 5mm
\begin{tabular}{ll|lll|lll}
\multicolumn{2}{c|}{} & \multicolumn{3}{c|}{$\beta$} & \multicolumn{3}{c}{$\beta$} \\
\multicolumn{2}{c|}{} & 1.0 & 2.0 & 3.0 & 1.0 & 2.0 & 3.0 \\
\hline\hline
 & $5.0$ & {\bf 1.31} & 1.15 & 1.11 & 1.24 & 1.13 & 1.10 \\
$\alpha$ & $6.0$ & 1.30 & 1.14 & 1.14 & {\bf 1.25} & 1.12 & 1.14 \\
 & $7.0$ & 1.27 & 1.13 & 1.16 & 1.23 & 1.12 & 1.17 \\
\hline
\multicolumn{2}{c|}{} & \multicolumn{3}{c|}{fix. RT} & \multicolumn{3}{c}{var. RT} \\
\end{tabular}

\caption{\label{tab:FDR_RT_risk} Ratio of binary classification risks with respect to the lowest attainable (oracle) risk for FDR control at different levels (top) and for the RT procedure with fixed ($K_n = 5\,000$) and variable ($\kappa_n = 5\,000$) window width (bottom), averaged over $100$ simulated datasets.}
\end{table}

It can be seen that both RT approaches perform in general better than FDR control through the BH procedure, with a slight advantage to the varying window extension. Most importantly, the classification risks they attain is never more than $1.31$ and $1.25$ times the oracle risk for the fixed and varying window version, respectively. In contrast to these near optimal performances, whatever the chosen level of FDR control, the BH procedure always performs poorly for at least one model, with an average classification risk that rises as high as $6.02$ times the optimal one in the worst case.

These results suggest that the RT approach, due to its adaptive nature, is indeed more stable than error rate control techniques, that depend on the choice of a false detection level, as we had anticipated. Moreover, the excellent performance of the RT methods, which attained near optimal risks on the studied datasets, is very encouraging for this approach.

\subsection{Results with unknown null distribution}

To illustrate the unknown distribution case, we simulated $n = 1\, 000$ observations $Y_i,$ among which $n_1 = 100$ where sampled from the $\mathcal N(\mu, \sigma^2)$ distribution with $\mu > 0$ and represented the significant terms, and $n - n_1 = 900$ where sampled from the $\mathcal N(0, 1)$ distribution and represented the null terms. We used less observations than in the known distribution case because the methods used in the present case are much more computer-intensive.

We implemented an EM algorithm to estimate a two-class Gaussian mixture model (GMM) from the data, with one zero-mean class to model the null data. As is often the case with iterative algorithms, providing initial values for the model parameters was the main problem we encountered. We found an efficient strategy for doing so, taking advantage of the fact that the negative data contained mostly null terms, and could provide a good initial guess for the null distribution variance and mixture weight. Details of the algorithm are given in Appendix~\ref{app:EM}.

Table~\ref{tab:GMM_RT_risk}, top shows the average ratios of the classification risks obtained by the different methods with respect to the oracle risk, over $100$ simulated datasets and for different choices of the Gaussian distribution parameters for the significant terms.

\begin{table}
\centering
\begin{tabular}{ll|lll|lll|lll}
\multicolumn{2}{c|}{} & \multicolumn{3}{c|}{$\sigma$} 
& \multicolumn{3}{c|}{$\sigma$} & \multicolumn{3}{c}{$\sigma$} \\
\multicolumn{2}{c|}{} & 1.0 & 2.0 & 3.0 & 1.0 & 2.0 & 3.0 & 1.0 & 2.0 & 3.0 \\
\hline\hline
 & $1.0$ 
 & 1.03 & 1.03 & 1.08
 & 1.03 & 1.06 & 1.02
 & 1.03 & 1.06 & 1.03 \\
$\mu$ & $2.0$ 
 & 1.06 & 1.03 & 1.04
 & 1.32 & 1.13 & 1.05
 & 1.30 & 1.12 & 1.05 \\
 & $3.0$ 
 & {\bf 1.11} & 1.06 & 1.04
 & {\bf 1.60} & 1.19 & 1.08
 & {\bf 1.55} & 1.18 & 1.08 \\
\hline
\multicolumn{2}{c|}{} & \multicolumn{3}{c|}{GMM fit}
& \multicolumn{3}{c|}{fix. RT} & \multicolumn{3}{c}{var. RT} \\
\end{tabular}
\vskip 5mm
\begin{tabular}{l|l|l}
{\bf 4.01} & {\bf 2.03} & {\bf 1.89} \\
\hline
GMM fit & fix. RT & var. RT\\
\end{tabular}

\caption{\label{tab:GMM_RT_risk} Top: Ratio of binary classification risks over oracle risk for model-based clustering (left) and for the RT procedure with fixed ($K_n = 5\,00$) and variable ($\kappa_n = 5\,00$) window width (middle and right), averaged over $100$ simulated datasets. Bottom: Results for bimodal non-null data (averaged over $100$ simulated datasets).}
\end{table}

All methods performed satisfyingly, yielding close to optimal risks, with model-based clustering performing slightly better than the RT methods. This comes as a little surprise, since in this case the former approach had several advantages: it was based on a parametric model for the significant terms that was precisely the one used to simulate the data, and it explicitly minimized the binary classification risk. In contrast, the RT approach requires no parametric assumption on the distribution of significant terms, and does not explicitly minimize a classification risk, but nevertheless gave good results.

Furthermore, performances of the model-based method can deteriorate when it is based on the wrong assumptions. To illustrate this, we simulated $n = 5\, 000$ observations $Y_i$ among which $n_1 = 950$ where sampled from the $\mathcal N(3, 1)$ distribution, $n_2 = 50$ from the $\mathcal N(20, 1)$ distribution, and the remaining $4\, 000$ from the $\mathcal N(0, 1)$ distribution and represented the null data. Consequently, the significant terms had a bimodal distribution. Most of these terms where next to the null mode, and a small number where next to a more distant mode.

This way, we hoped to trick the mixture model, which assumed a unimodal distribution for the significant terms, in detecting only the distant mode, while merging the other one with the null distribution. This is exactly what happened, as can be seen in Table~\ref{tab:GMM_RT_risk}, bottom: the mixture-model fit performs significantly worse than the RT approach in this case, the later maintaining a reasonable, though also degraded, classification risk.

Of course it can be argued that such a dataset does not represent a realistic situation; our point here is simply to illustrate the increased robustness of RT due to the fact that it requires no assumptions other than a noise model. It can also be discussed that an alternative to the simplistic two-class GMM used here would be to allow a variable number of classes, combined with a model selection framework \cite{Hastie01,Massart03,Efron04}. However, implementing such complex strategies would be non-trivial, especially concerning the algorithm's initialization. This last issue could be addressed for instance by using stochastic extensions of the EM, such as the stochastic averaging EM (SAEM) \cite{Delyon99}, in order to reduce dependence on initial values. In contrast to such sophisticated strategies, the simplicity of the RT approach, which requires minimal implementation and virtually no tuning, appears as a key advantage in practice, especially in view of the good performances suggested by this study.

\section{Application to fMRI data analysis}\label{sec:fMRI}

We now apply the random threshold approach to functional magnetic resonance imaging (fMRI) data analysis. fMRI is a modality of {\em in vivo} brain imaging that allows to measure the variations of cerebral blood oxygen levels induced by the neural activity of a subject lying inside a MRI scanner and submitted to a series of stimuli. A sequence of three-dimensional (3D) images of the brain is thus acquired, measuring over time a vascular effect of neural activity known as the blood oxygenation level dependent (BOLD) effect.  From the time series recorded in each voxel, and the occurrence times for each stimulus, one may compute an estimate of the BOLD effect of the subject in response to any given stimulus, and more generally to any difference or combination of stimuli (contrast) \cite{Friston97,Worsley02}.

Thus, the fMRI data for one subject generally consists in a spatial map of $z$-scores $(Y_1, \ldots, Y_p),$ where $p$ is the number of voxel in the search volume (which can be as high as $100\, 000$), and $Y_i$ the estimated BOLD effect at voxel~$i.$ This map of measures of cerebral activity, also termed {\em activation map}, is plagued by several sources of uncertainty: the natural variability of brain activity, and the estimation noise induced by the MRI scanner. Thus, model~(\ref{eq:obs_model}) provides a good representation of the activation map $(Y_1, \ldots, Y_p),$ with significant terms corresponding to brain regions involved in the task under study.

In a typical fMRI study however, not one but several subjects are recruited from a population of interest, and scanned while submitted to the same series of stimuli. Activation maps associated with a given contrast are obtained for each subject, as described above, and used as input data for inference at the between-subject level, where the goal is to evidence a general brain activity pattern. Mass univariate, or voxel-based, detection \cite{Friston97} is to date the most widely used approach to address this question. It starts with normalizing individual images onto a common brain template using nonrigid image registration. Next, a $t$-statistic is computed in each voxel to locally assess mean group effects.

In both single-subject and multi-subject fMRI data analysis, the problem of activation detection can be formulated statistically as that of detecting non-zero means among a collection of observations. The most common approach consists in thresholding a statistical map of brain activity \cite{Friston97}. Multiple-testing techniques are widely used \cite{Nichols03,Pacifico04}, as well as mixture-models. The Gamma-Gaussian mixture model (GGM) is most often used in this context \cite{Beckmann03b}. It uses a Gaussian distribution for null, or inactivated, data, a Gamma distribution for activated data, and a negative Gamma distribution for deactivated data.

As these methods suffer from certain limitations, as discussed in the previous Sections, the RT appears as an appealing alternative in this context. Hence we decided to compare the regions detected by the different approaches, to see if RT succeeded in recovering regions known to be involved in certain well-studied cognitive tasks.

\subsection{Data acquisition and preprocessings}

We used a real fMRI dataset from the Localizer database~\cite{Pinel07}, involving a cohort of 38 right-handed subjects, and acquired as follows. The participants were presented with a series of stimuli or were engaged in tasks such as passive viewing of horizontal or vertical checkerboards, left or right click after audio or video instruction, computation (subtraction) after video or audio instruction, sentence listening and reading. Events occurred randomly in time (mean inter stimulus interval: 3s), with ten occurrences per event type, and ten event types in total.

Functional images were acquired on a General Electric Signa 1.5T scanner using an Echo Planar Imaging sequence . Each volume consisted of $34$ $64 \times 64$ $3$~mm-thick axial contiguous slices. A session comprised $130$ scans. Anatomical T1 weighted images were acquired on the same scanner, with a spatial resolution of $1\times1\times1.2$~mm$^3$. Finally, the cognitive performance of the subjects was checked using a battery of syntactic and computation tasks.

Single-subject analyses were conducted using SPM5 ({\tt http://www.fil.ion.ucl.ac.uk}). Data were submitted successively to motion correction, slice timing and normalization to the~MNI template. For each subject, BOLD contrast images were obtained from a fixed-effect analysis on all sessions. Group analyses were restricted to the intersection of all subjects' whole-brain masks, comprising $43\, 367$~voxels.

We considered the $t$-score maps computed for different contrasts of experimental conditions. These were first converted to $z$-score maps, to obtain approximatively Gaussian statistics in inactivated voxels. Using these maps as input data, we then compared the detection thresholds obtained by Gamma-Gaussian mixture modeling (GGM), fixed-window random thresholding and the varying-window extension, also using the unknown variance extension in both cases (see Section~\ref{sec:unknown}). For simplicity, we only present here the results obtained for a fixed window equal to $K_n = 15\,000.$

\subsection{Individual subject activation map}

\begin{figure}
\centering
\begin{minipage}{1.0\textwidth}
\includegraphics[width=0.23\textwidth]{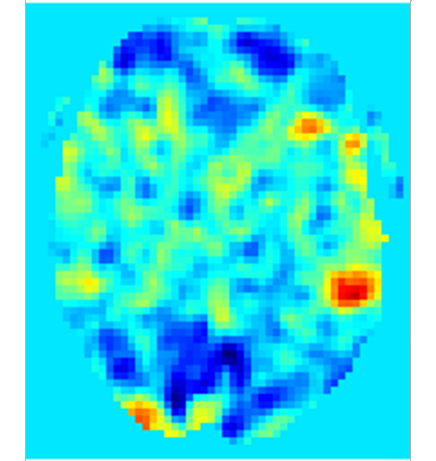}\hfill
\includegraphics[width=0.23\textwidth]{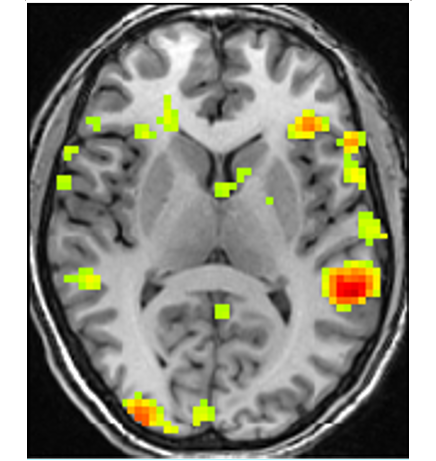}\hfill
\includegraphics[width=0.23\textwidth]{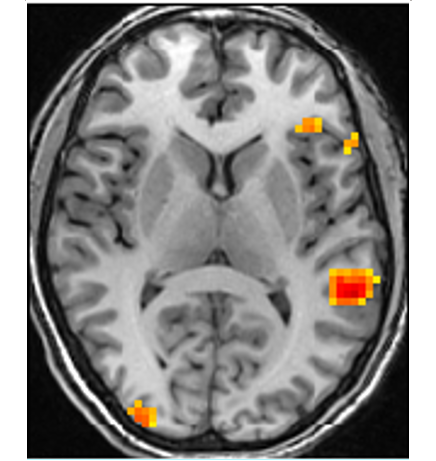}\hfill
\includegraphics[width=0.23\textwidth]{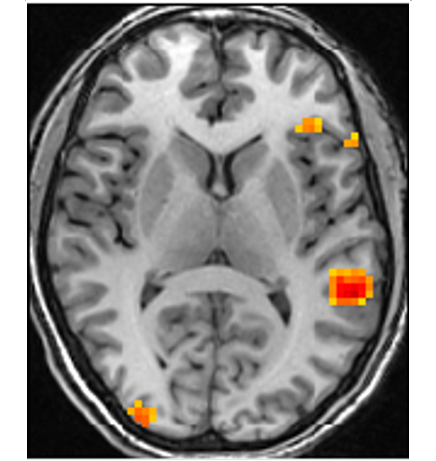}
\end{minipage}
\caption{\label{fig:intra_maps}  Axial slice from a $z$-score map for the "sentence--checkerboard" contrast, using a temperature palette for the $z$-score values. From left to right: Unthresholded, thresholded by GGM model fit, varying-window and fixed-window ($K_n = 15\, 000$) random thresholding. Detected activations are shown against the subject's anatomical image.}
\end{figure}

Our first illustration concerns the activation map of a single subject, for the ``sentence--checkerboard'' contrast. This contrast subtracts the effect of viewing horizontal and vertical checkerboards from that of reading video instructions, thus allowing to detect brain regions specifically implicated in the reading task.

Figure~\ref{fig:intra_maps}, left, shows an axial slice from the $z$-score map before thresholding. Activations are clearly seen in Wernicke's and Broca's areas (right and upper right), which are known to be involved in language processing (see \cite{Price00}, for instance). The detection threshold found by GGM fit for the $z$-score map ($2.03$) is much lower than those found by the random threshold procedure, both with a varying window ($3.19$) and a fixed window ($3.33$).

The random thresholds with fixed and variable windows yield very similar activation maps in this case, which seem to capture the activated regions seen in the raw map. In contrast, the much lower threshold found by mixture modeling detects several smaller clusters, some of which may be false positives.

\subsection{Group activation map}

In this second example, we consider a group activation map, specifically a map of $t$-statistics computed from the
individual contrast maps of 15 subjects, thus enabling to infer regions of positive mean effects in the parent population.
Our choice of limiting the number of subjects, rather than using the whole cohort, was driven by the fact that many fMRI
studies are conducted on groups of less then 20 subjects.

We report results for the ``calculation--sentences'' contrast, which
subtracts activations due to reading or hearing instructions from the
overall activations detected during the mental calculation tasks. This
contrast may thus reveal regions that are specifically involved in the
processing of numbers.

Figure~\ref{fig:intra_maps}, left, shows an axial slice from the activation map before thresholding, with clear activations in the bilateral anterior cingulate (upper middle), bilateral parietal (lower left and right) and right precentral (upper right) regions, all known to be involved in number processing \cite{Pinel07}.

Though sorted in the same order as previously, the varying window random threshold ($2.49$) is now roughly at equal distances from the threshold found by GGM modeling ($1.79$) and the fixed window random threshold ($3.06$).

The three methods detected activations in the regions described above, though the fixed window random threshold seemed to miss some
activations, and the GGM approach further detected smaller clusters, some of which may be false positives.

Of course one cannot conclude from these examples only that RT is `better' at detecting activations than GGM fit. However, the varying window extension successfully detected regions known to be involved in the two cognitive tasks considered here, while avoiding isolated peaks in other regions, which may be part of the background noise. These results suggest that the RT succeeded in capturing only the active regions, while the GGM approach seemed to detect spurious activations.

\begin{figure}
\centering
\begin{minipage}{1.0\textwidth}
\includegraphics[width=0.23\textwidth]{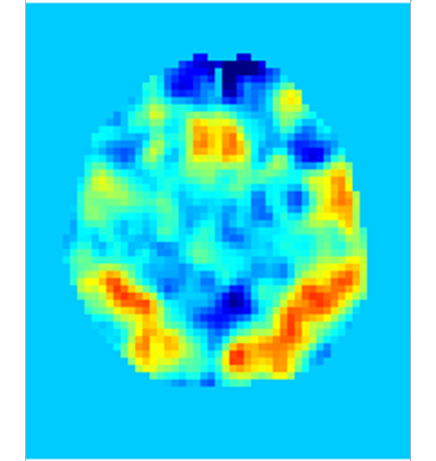}\hfill
\includegraphics[width=0.23\textwidth]{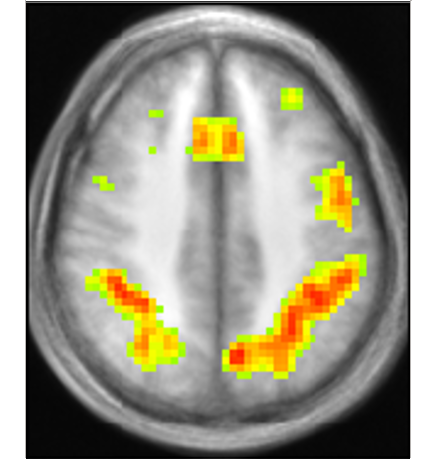}\hfill
\includegraphics[width=0.23\textwidth]{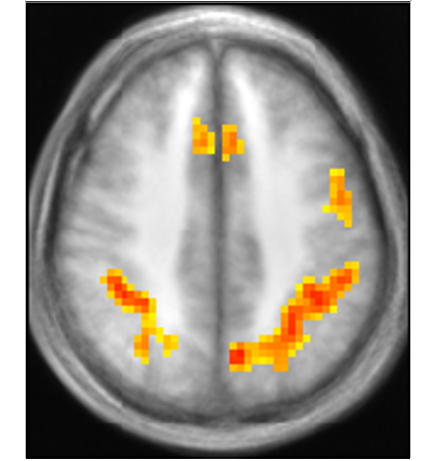}\hfill
\includegraphics[width=0.23\textwidth]{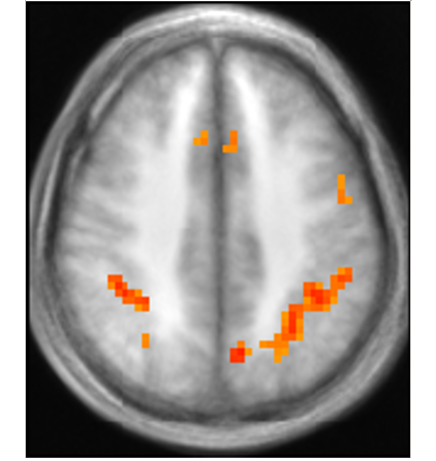}
\end{minipage}
\caption{\label{fig:inter_maps}  Axial slice from the group activation $z$-score map for the "calculation--sentence" contrast, using a temperature palette for the $z$-score values. From left to right: Unthresholded, thresholded by GGM model fit, varying-window and fixed-window random thresholding. Detected activations are shown against the mean anatomical image of all subjects.}
\end{figure}

\section{Conclusion}\label{sec:conclusion}

In this paper, we have introduced a simple modification to the random threshold (RT) procedure proposed in \cite{Lavielle07}, to obtain an entirely unsupervised procedure for recovering non null mean terms from a collection of independent random observations, based solely on a parametric model of the null terms. Our modification, which requires no prior tuning, conserves the consistency properties of the original procedure.

We have implemented all the different versions of the random threshold method in a Python package. This was integrated to the Neuroimaging in Python (NIPY) open-source library, freely downloadable from \url{http://nipy.sourceforge.net}.

We validated this approach through extensive numerical experiments, and showed that both the original procedure, based on a fixed window-width, and our extension, which uses a variable window, compare favorably to both multiple testing procedures, and model-based clustering, in terms of the binary classification risk, with a slight advantage to our varying window extension. On the vast majority of simulated datasets, the risks achieved where close to the lowest achievable (oracle) risk, whereas each of the other approaches behaved poorly in at least one case.

Thus RT appears as a very promising method for non-ordered model selection whenever no parametric assumptions are available concerning the data distribution. Such methods are needed in many application domains, as we have illustrated in the case of activation detection for fMRI data analysis.

The good classification performances of RT evidenced empirically in our simulations suggest that a promising direction for future research would be to study its properties in the mixture-model setting, and especially its large-sample behaviour. An interesting question to answer would be whether the random threshold converges to a certain limit when the SNR remains constant, and if so, how does this limit compares to the oracle threshold.

\appendix

\section{Proof of the Theorem \ref{theo:consistency} for the varying window extension}\label{app:proof}

Following \cite{Lavielle07}, we first recall some notations. Set $U_i = Y_i$ for $i \in I_{k_n^\star}$ and $v_i = Y_i$ for $i \notin I_{k_n^\star};$ notice that $(v_i)$ is a sample from the distribution $F_{\epsilon}.$ Let $(u_{(i)})_{1\leq i\leq k_n^\star}$ and $(v_{(i)})_{1\leq i\leq n - k_n^\star}$ be the sequences $(|U_i|)$ and $(|v_i|)$ in decreasing order. Let $\Omega_n$ be the subset of $\Omega$ where $v_{(1)} < \alpha_n / 2$ and $u_{(k_n^\star)} > \alpha_n / 2.$

A first lemma in \cite{Lavielle07} shows that $P(\Omega_n) \to 1,$ \emph{i.e.}, the collections $(u_{(i)})$ and $(v_{(i)})$ are stochastically in order with high probability. The proof can then be restricted to $\Omega_n.$

Now, let $\mathbb{E}_k(T_{k,j})$ and $Q_{k,j}$ be defined as in Equation~(\ref{eq:procedure}). Using Proposition~\ref{theo:cond_mean}, we have:

\begin{eqnarray}
\mathbb{E}_k(T_{k,j}) & = & j ( 1 + \sum_{i = j + 1}^{n - k} 1 / i ); \nonumber\\
Q_{k,j} & = & \frac{\mathbb{E}_k(T_{k,j})}{\mathbb{E}_k(T_{k,n - k})} T_{k,n - k} \nonumber\\
 & = & B_{k,j,n} T_{k,n - k}.\nonumber
\end{eqnarray}

Also, let $a_i = \mathbb{E}_0(Z_{(i)}) = \sum_{\ell = i}^n 1/\ell.$ Equation~(\ref{eq:consistency}) can be shown separately for $k > k_n^\star$ and $k < k_n^\star.$ Since the two cases are treated similarly, we will restrict ourselves here to the case $k > k_n^\star.$ On $\Omega_n:$

\begin{eqnarray}
T_{k,j} - Q_{k,j} 
& = & T_{k,j} - B_{k,j,n} T_{k,n - k} \nonumber\\
& = & \left( T_{k,j} - \mathbb{E}_{k_n^\star}(T_{k,j}) \right) - B_{k,j,n} \left( T_{k,n - k} - \mathbb{E}_{k_n^\star}(T_{k,n - k}) \right)\nonumber\\
&   & + \mathbb{E}_{k_n^\star}(T_{k,j}) - B_{k,j,n} \mathbb{E}_{k_n^\star}(T_{k,n - k}) \nonumber\\
 & = & R_{k,j} + S_{k,j}.\nonumber
\end{eqnarray}

Thus $T_{k,j} - Q_{k,j}$ is decomposed into a random part $R_{k,j}$ and a deterministic part $S_{k,j}.$ Over $\Omega_n,$ $R_{k,j}$ is a function of $v_{(k)}, \ldots, v_{(n - k_n^\star)}.$ Before going further, we now recall the following result:

Let $Z_{(1)} \geq \ldots \geq Z_{(n)}$ be an ordered sequence of independent $Exp(1)$ random variables. For $1 \leq j \leq n,$ let $T_j = \sum_{i = 1}^j Z_{(i)}.$ Introduce for $t \in [0,1]$ the random process $d_n(t) = T_{[nt]} - \mathbb{E}(T_{[nt]} | T_n).$ Then it is shown in \cite{Lavielle07} that $\frac{1}{\sqrt{n}}d_n(t),$ as a process indexed on $t \in [0,1],$ converges in distribution to a certain zero mean Gaussian process $\Delta.$

To use this result, let $k = [tn]$ and $j = [sn],$ for $0 < t < 1 - c$ and $0 < s < 1 - t^{\star} - t,$ for $c$ in [{\bf AF3}]. Then $\frac{1}{\sqrt{n-k}}(T_{k,j} - Q_{k,j})\mathbf{1}_{\Omega_n} = \frac{1}{\sqrt{n-k}}(T_{[tn],[sn]} - Q_{[tn],[sn]})\mathbf{1}_{\Omega_n},$  as a process indexed by $(t,s) \in (0,1)^2,$ converges in distribution to the zero-mean Gaussian process:

$$
\Gamma_{t,s} = \sqrt{\frac{1 - t^{\star}}{1 - t}}\left[ \Delta \left( \frac{t + s - t^{\star}}{1 - t^{\star}} \right) - \Delta \left( \frac{t - t^{\star}}{1 - t^{\star}} \right) \right].
$$

similarly, $\frac{1}{\sqrt{n - k}}B_{k,j,n} \mathbb{E}_{k_n^\star}(T_{k,n - k})\mathbf{1}_{\Omega_n}$ converges in distribution to another zero-mean Gaussian process, and so does their sum, $\frac{1}{\sqrt{n - k}}R_{k,j}\mathbf{1}_{\Omega_n}.$

On the other hand, 

\begin{eqnarray}\label{eq:cond_mean}
S_{k,j} & = & \sum_{i = 1}^{k - k_n^\star} (a_{i+j} - a_i + B_{k,j,n}(a_{i+n-k} - a_i)),
\end{eqnarray}

so that there exists a constant $\gamma > 0,$ which depends on $c$ in [{\bf AF3}], such that for all $n \geq 1,\ k_n^\star < k \leq n - K_n,$ we have $\sup_{1 \leq j \leq n - k} |S_{k,j}| \geq \gamma (k - k_n^\star).$ Finally we use the following inequality:

\begin{eqnarray}
\mathbb{P}_{k_n^\star} (\hat k_n - k_n^\star > n u_n) & \leq & \mathbb{P}( \eta_{k_n^\star} > \inf_{k - k_n^\star > n u_n} \eta_k ). \nonumber
\end{eqnarray}

From Equation~(\ref{eq:cond_mean}), $S_{k_n^\star,j} = 0,$ hence it follows that:
\begin{eqnarray}
\sqrt{n - k_n^\star}\, \eta_{k_n^\star} 
&=& 
\sup_{1 \leq j \leq n - k_n^\star} |R_{k_n^\star,j} + S_{k_n^\star,j}| \nonumber\\
&=& 
\sup_{1 \leq j \leq n - k_n^\star} |R_{k_n^\star,j}|  \nonumber\\
&\leq& \sup_{k \geq k_n^\star} \sup_{1 \leq j \leq n - k} |R_{k,j}|.\nonumber
\end{eqnarray}
On the other hand,
\begin{eqnarray}
\sqrt{n - k}\inf_{k - k_n^\star > n u_n} \eta_k 
& = & \inf_{k - k_n^\star > n u_n} \sup_{1 \leq j \leq n - k} |R_{k,j} + S_{k,j}| \nonumber\\
&\geq& \inf_{k - k_n^\star > n u_n} \sup_{1 \leq j \leq n - k} |S_{k,j}| - \sup_{k \geq k_n^\star} \sup_{1 \leq j \leq n - k} |R_{k,j}|, \nonumber
\end{eqnarray}
so that we have:
\begin{eqnarray}
\mathbb{P}_{k_n^\star} (\hat k_n - k_n^\star > n u_n) & \leq & \mathbb{P}(C \sup_{k \geq k_n^\star} \sup_{1 \leq j \leq n - k} |R_{k,j}| \geq \inf_{k - k_n^\star > n u_n} \sup_{1 \leq j \leq n - k} |S_{k,j}|) + \mathbb{P}(\Omega_n^c) \nonumber\\
                                                         & \leq & \mathbb{P}(C \sup_{k \geq k_n^\star} \sup_{1 \leq j \leq n - k} |R_{k,j}| \geq \gamma n u_n) + \mathbb{P}(\Omega_n^c), \nonumber
\end{eqnarray}
where $C$ is a constant which depends on $c$ in [{\bf AF3}]. This last probability vanishes as $n$ goes to infinity, due to the weak convergence of $R_{k,j}\mathbf{1}_{\Omega_n} \square$

\section{Details of the EM algorithm for the two-class GMM with a zero-mean class}\label{app:EM}

We consider the following model:

\begin{eqnarray}
Y_i | Z_i=j &\stackrel{iid}{\sim}& \mathcal N(\mu_j, \sigma_j^2), \quad i = 1, \ldots, n,\quad j = 0, 1 \nonumber\\
Z_i &\stackrel{iid}{\sim}& \mathcal B(1, p_1),
\end{eqnarray}
where $\mu_0 = 0,$ and $p_j$ represents the proportion of data in class $j,$ so that the vector of model parameters is: $\theta = (p_0, \mu_1, \sigma_0, \sigma_1).$

Having initialized $\theta$ to $\theta^{(0)},$ the EM algorithm alternates the following steps:

\begin{itemize}
\item[] {\bf E-step.} Compute the conditional law of the indicator variable $Z_i$ at step $t,$ that is, the Bernoulli defined by:
\begin{eqnarray}
\mathbb P(Z_i = j | Y_i, \theta^{(t)}) &=& \frac{f(Y_i | Z_i = j, \theta^{(t)}) p_j^{(t)}}{\sum_j f(Y_i | Z_i = j, \theta^{(t)}) p_j^{(t)}}\nonumber\\
&:=& p_{ij}^{(t)}
\end{eqnarray}

\item[] {\bf M-step.} Update the estimate of model parameters by maximizing the conditional expectation of the complete log-likelihood, yielding:
\begin{eqnarray}
\theta^{(t+1)} &=& \arg\max_\theta \mathbb E [\sum_i \log f(Y_i | Z_i, \theta) | \bs Y, \theta^{(t)}],\nonumber
\end{eqnarray}
the expectation being taken with respect to the conditional distribution of the indicator variables $Z_i$ computed in the previous step. This yields:
\begin{eqnarray}
p_j^{(t + 1)} &=& \frac{\sum_i p_{ij}^{(t)}}{n}; \nonumber\\
\mu_1^{(t+1)} &=& \frac{\sum_i p_{ij}^{(t)} Y_i}{\sum_i p_{ij}^{(t)}}; \nonumber\\
{\sigma_j^2}^{(t+1)} &=& \frac{\sum_i p_{ij}^{(t)} (Y_i - \mu_j^{(t)})}{\sum_i p_{ij}^{(t)}}. \nonumber\\
\end{eqnarray}

\end{itemize}

Note that, throughout the iterations, $\mu_0^{(t)} \equiv 0.$

\paragraph{Initialization.} An initial guess for $\sigma_0^2$ is provided by the negative data, which consists mainly of null data:
\begin{eqnarray}
{\sigma_0^2}^{(0)} &=& \sharp\{Y_i < 0\}^{-1} \sum_{Y_i > 0} Y_i^2 \nonumber.
\end{eqnarray}

Then, we use a kernel estimate of the data density:

\begin{eqnarray}
\hat f(x) &=& \frac{1}{n} \sum_i \frac{K\left(\frac{x - Y_i}{h_n}\right)}{h_n},
\end{eqnarray}
for a symmetric, positive kernel $K,$ and a bandwidth $h_n.$ In pratice, we used the Gaussian kernel $K(x) = {2\pi}^{-1/2}e^{-x^2/2},$ and $h_n = \sqrt n.$

By identifying the null mode of the data density kernel estimate to the null component of the mixture model, we then obtained an initial guess for the mixture weights:

\begin{eqnarray}
p_0^{(0)} &=& \hat f(0) \sqrt{2\pi{\sigma_0^2}^{(0)}}.\nonumber
\end{eqnarray}

Finally, the conditional law of the indicator variables where approached by:
\begin{eqnarray}
p_{i1}^{(0)} &=& \min 
\left\{ 
1; \frac
{p_0^{(0)} \exp\{-(Y_i^2 / 2 {\sigma_0^2}^{(0)})\} }
{\sqrt{2\pi{\sigma_0^2}^{(0)}}}  \right\}.\nonumber
\end{eqnarray}
These initial guesses are used to derive initial model parameter values $\theta^{(1)},$ via the M-step described above, for $t=0.$

\end{document}